\documentclass[10pt,twocolumn]{IEEEtran}
\usepackage{amsmath,amsthm,amssymb,epsfig,subfigure,cite}
\usepackage{algorithm,algpseudocode}

\newtheorem{theorem}{\hspace*{1pc}Theorem}
\newtheorem{corollary}{\hspace*{1pc}Corollary}
\newtheorem{lemma}{\hspace*{1pc}Lemma}

\begin{document}

\bibliographystyle{unsrt}

\setlength{\parindent}{1pc}

\title{Loss Rate Inference in Multi-Sources and Multicast-Based General Topologies}
\author{Weiping~Zhu,\IEEEmembership{ member, IEEE}  \thanks{Weiping Zhu is with University of New South Wales, Australia}}
\date{}
\maketitle

\begin{abstract}
 Loss tomography has received considerable attention in
recent years and a number of estimators have been proposed.
Unfortunately, almost all of them are devoted to
 the tree topology despite the general topology is more common in practice. In addition,  most of the works presented in the literature
 rely on iterative
approximation to search for the maximum of a likelihood function
formed from observations, which have been known neither scalable nor
efficient. In contrast to the tree topology, there is few paper
dedicated to the general topology because of the lack of understanding
the impacts created by the probes sent by different sources. We in
this paper present the analytical results obtained recently for the
general topology that show the correlation created by the probes sent
by multiple sources to a node located in an intersection of multiple
trees. The correlation is expressed by a set of polynomials of the
pass rates of the paths connecting the sources to the node. In
addition to the expression, a closed form solution is proposed to
obtain the MLE of the pass rates of the paths connecting the sources
to the node. Then, two strategies are proposed to estimate the loss
rate of a link for the general topology: one is path-based and the
other is link-based, depending on whether we need to obtain the pass
rate of a path first. The two strategies are compared in the context
of the general topology that shows each has its advantages and the
link-based one is more general. Apart from proving the estimates
obtained are the MLEs, we prove the estimator presented here has the
optimal asymptotic property.
\end{abstract}


\section{Introduction}

Network tomography was proposed in \cite{YV96} that suggests the use
of end-to-end measurement and statistical inference to estimate
network characteristics of a large network. Since then, a considerable
amount of works has been done in this area and the works reported so
far have covered almost all of the characteristics that were
previously obtained by direct measurement for a small network. The
works can be classified into loss tomography \cite{Duffield2002},
delay tomography \cite{LY03}, \cite{TCN03}, \cite{PDHT02},
\cite{SH03}, \cite{LGN06}, loss pattern tomography \cite{ADV07},
source-destination traffic matrix \cite{LY06} and shared congestion
flows \cite{RKT02}. Despite such a overwhelming enthusiasm and a
wealth of publications, most of the works in this area are at the
preliminary stage, and most of the estimators or algorithms proposed
for estimations aim at proof of concept while the efficiency and
scalability of them are very much overlooked. For instance, loss
tomography has been studied for a decade and a number of algorithms
have been published to estimate link-level loss rates from end-to-end
measurement. Unfortunately, almost all of the algorithms are limited
into the tree topology and most of them resort on iterative
approximation. To overcome this problem, we in this paper propose a
number of maximum likelihood estimators to estimate link-level loss
rate of a general network, all of them are in closed form.

The topologies used to connect sources to receivers can be divided
into two classes: tree and general, which leads to two types of
estimators, one for a topology. As stated, the tree one has received
much more attention than its counterpart. Almost all of the estimators
proposed previously target the tree topology because of the simplicity
of the correlation embedded into the observations of the receivers
attached to a tree topology.  Despite the large amount of works, there
are only a few of them providing analytical solutions as \cite{CDHT99}
\cite{DHPT06} \cite{Zhu06-2}. That leads to the use of iterative
procedures to approximate the root of a likelihood function although
knowing such a  procedure is neither efficient nor scalable
\cite{DLR77}. In contrast to the tree topology, there has been little
systematically study in both theory and experiments for the general
topology despite the general topology is more common than the tree one
in practice. As a result, there are only  a handful of estimators that
have been proposed for the general topology. To improve this, we in
this paper provide our findings about the impact of multiple sources
on loss estimation that lead to a number of maximum likelihood
estimators for the general topology, including an analytical solution,
and two divide-and-conquer ones. All of them completely eliminate
iterative procedure involved in the estimation. Compared with the
estimators proposed previously for the general topology
\cite{CDHT99},\cite{BDPT}, the estimators proposed here not only
perform better, but also extend our understanding of loss tomography
from the tree topology to the general one.  In fact, the results
presented in this paper generalize those obtained from the tree
topology since the tree is a special case of the general topology.
Further, the results may be extended into other tomographies, such as
delay, in the general topology.

\subsection{Related Works}
Multicast Inference of Network Characters (MINC) is the pioneer of
using multicast probes to create correlated observations at the
receivers of the tree topology, where a Bernoulli model is used to
model the losses occurred on a link. Using this model, the authors of
\cite{CDHT99} derive a direct expression of the pass rate of a path
connecting the source to an internal node. The expression is in the
form of a  polynomial \cite{CDHT99}, \cite{CDMT99}, \cite{CDMT99a}.
 To ease the concern of unsalability of
using numeric method to solve higher degree polynomials $(
>5 )$, Duffield {\it et. al.}  propose an explicit estimator in \cite{DHPT06} for the tree topology. Although the estimator has the same asymptotic variance as the
MLE to first order, it is not an MLE. When $n < \infty$, the estimate
obtained by the estimator can be very different from an MLE. In
contrast, \cite{CDHT99} proposed an iterative approach to search for
the maximum of the general topology. Later, Bu {\it et. al.} attempted
to use the same model as the one used in \cite{CDHT99} for the general
topology, but could not derive a direct expression as its predecessor
\cite{BDPT}. Without analytic results, the authors of \cite{BDPT} then
resort on iterative procedures, e.g. the
 EM, to approximate the
 maximum of the likelihood function. In addition, a minimum variance weighted average (MVWA) estimator is proposed as an alternative in comparison with the EM
 one.
 The MVWA in
 fact is a variant of the minimum chi-square estimator
\cite{Rao80}. The experimental results
 confirm
this problem \cite{BDPT}, where the EM outperforms the MVWA when the
sample size is small. Nevertheless, there is no proof that the
solution obtained by the EM algorithm is the MLE since the iterative
procedures, e.g. EM algorithm, can converge to a local maximum unless
the solution space is proved to be strictly concave.

Considering the unavailability of multicast in some networks, Harfoush
{\it et al.} and Coates {\it et al.} independently proposed the use of
the unicast-based multicast to discover link-level characteristics
\cite{HBB00}, \cite{CN00}, where Coates {\it et al.} also suggested
the use of the EM algorithm to estimate link-level loss rates that may
not be scalable as previously stated.

To improve the scalability of estimation, Zhu and Geng propose a
bottom up algorithm in \cite{ZG04-3}. The algorithm is independent of
topology, i.e. it can be applied to the general topology as well
\cite{ZG05}. The estimator, rather than estimating the loss rates of
all links together as an iterative procedure, adopts a step by step
approach to estimate the loss rate of a link one at a time and from
bottom up. At each step the algorithm uses a closed form formula to
estimate the loss rate of a link. Despite the effectiveness,
scalability and extensibility to the general topology,
 the estimator is not the MLE as the one presented in \cite{DHPT06}
 because the statistics used in estimation are not sufficient.
 Recently, the authors of \cite{ZD09} use a divide-and-conquer approach to tackle the general
 topology,
 which divides a general network into a number of independent trees at the roots of intersections. It uses a fixed point algorithm to
 estimate the number of probes reaching the roots of intersections. The fixed point algorithm
 is based on an iterative procedure to achieve its goal, although experimental studies show
 it has a fast convergence rate since the solution space is strictly
concave.

\subsection{Contribution and Paper Organization}
As stated,  there has been a lack of understanding the impact of
probes sent from multiple sources to a link or a node in the general
topology. In addition, iterative procedures are still used in all
estimators proposed for the general topology \cite{ZD09}. We are
aiming to improve the unsatisfied situation and present the latest
findings in this paper that contribute to loss tomography in the
following respects:

\begin{enumerate}
\item we derive a set of polynomials to express the pass rate of the paths connecting a source to a node in the general topology. The derivation shows
the pass rate of a path in the general topology is no longer
independent from each other in contrast to that in a tree topology.
This finding generalizes the results of \cite{CDHT99} and is
applicable to various topologies.
\item we derive a closed form MLE solution to the set of the polynomials of related paths. The solution shows that the MLE
considers the probes sent from all related sources that generalizes
the result obtained from the tree topology.
\item we present a divide-and-conquer strategy to break up a general network into a number of independent trees to avoid the necessity of
solving a large number of polynomials. Two algorithms are proposed
subsequently to the independent trees, one for the descendants of a
decomposing point, the other for the ancestors of the decomposing
point. The algorithms previously developed for the tree topology can
be applied to the former with little help while the latter requires
its own algorithm.
\item we propose two estimators: link-based and
path-based, for the ancestors of a decomposing point. The result shows
each of the estimators has its own advantages and disadvantages. The
link-based one is more general than the path-based one.
\item we clarify some misconceptions about using multiple sources, in
particular to those focusing on the complexity created by multiple
sources in estimation but ignoring the benefits brought by the
multiple sources on  data consistency, variance reduction, and
convergence rate.
\end{enumerate}

The rest of the paper is organized as follows. In Section 2 we present
the essential background, including the notations and statistics used
in this paper. In Section 3, we present our latest findings and
contributions to the loss inference of the general topology. Section 4
provides the detail description of the three estimators, plus some
asymptotical properties of the estimator proposed in this paper.
Section 5 is devoted to the statistical properties of the estimators
presented in this paper. Simulations are presented in Section 6 that
shows the estimates of overlapped links are converged quicker than
other links. The last section is devoted to concluding remark.

\section{Notations and Statistics} \label{section2}
A general topology differs from a tree one in a number of aspects. The
most important one is the use of multiple trees to cover a general
network, which leads some parts of the network covered more than once
by the trees. Those parts are called intersection areas, or simply
intersections. If a source is assigned to the root of a tree to send
probes to the receivers attached to the tree, the receivers attached
to an intersection can observe the probes sent from multiple sources.
Then, the question is whether we need to consider the probes sent by
different sources in estimation. If not, the estimate obtained from a
source can be different from that obtained from another source and the
difference can be noticeable unless the number of probes sent by each
source goes to infinite that is not appropriate in practice. If yes,
we must determine the contribution of each source on the estimate. In
addition, the 1-to-1 mapping between nodes and links held in the tree
topology no longer exists in the general topology since some nodes can
have more than one parents.

\subsection{Notation}\label{treenotation}

A multicast tree or subtree used to connect the source to receivers is
slightly different from an ordinary tree at
 the root  that has only a single child. The link connecting the root node to the child is called the root link of the
 tree. Let ${\cal N}$ denote a general
network consisting of $k$ trees, where $S=\{s^1, s^2, \cdot\cdot,
s^k\}$ denote the sources attached to the trees. In addition, $V$ and
$E$ denote the nodes and links of ${\cal N}$. Each node is assigned a
unique number, so is a link. $|V|$ and $|E|$ denote the number of
nodes and the number of link of the network, respectively. The two
nodes connected by a link are called the parent node and the child
node of the link, where the parent receives probes from its parent and
forward them to its children. In addition, $p(s, i)$ denotes the
parent link of link $i$ toward source $s$. The nodes that have no
parent(s) are the roots of the multicast trees. Each of the multicast
trees and the subtrees is named after the number assigned to its root
link, where $T(i), i \in E$ denotes the subtree with link $i$ as its
root link. Further, let $R(s), s \in S$ denote the receivers attached
to the multicast tree rooted at $s$ and let $Rs(i), i \in E$ denotes
the receivers attached to the multicast subtree rooted at link $i$.
The largest intersection between two trees is called the {\it
intersection} of the trees and the root of an intersection is called
the {\it joint node} of the intersection. Note that an intersection is
an ordinary tree but a multicast one. We use $J$ to denote all joint
nodes and use $S(j), j\in V \setminus S$ to refer the sources that
send probes to node $j$. If $f_1^j(i)$, simply $f^j(i)$ later, is used
to denote the parent of node $i$ on the way to source $s^j$ and
$f_l^j(i)$ to denote the ancestor that is $l$ hops away from node $i$
on the path to source $s^j$, we have $f_k^j(i)=f^j(f_{k-1}^j(i))$.
Further, let $a(j,i)=\{f^j(i), f_2^j(i),\cdot\cdot, f_k^j(i)\}$, where
$f_{k+1}^j(i)=s^j$, denote the ancestors of node $i$ to $s^j$. Let
$a(i)=\{a(j,i), j \in S(i)\}$ denote all the ancestors of node $i$. In
addition, except for leaf nodes each node has a number of children,
where $d_i$ denotes the children of node $i$ and $|d_i|$ denotes the
number of children of node $i$.

 If $n^s, s \in S$ denotes the
number of probes sent by source $s$, each probe $o=1,...., n^s$ gives
rise of an independent realization $X^s(o)$ of the probe process $X$,
$X_k^s(o)=1, k\in E$ if probe $o$ passing link $k$; otherwise
$X_k^s(o)=0$. The observation of $\Omega=\{\Omega_s, s \in S\},
\Omega_s=(X^s(o)), {o=1,2,...,n^s}$ comprise the data set for
inference. In addition, let $Y_k^s(i), {i=1,2,...,n^s}, k \in E, s \in
S$ denote the state of link $k$ obtained by examining $\Omega_s$ for
probe $i$ to see whether it reaches at least one of $Rs(k)$.
$Y_k^s(i)=1$ if the probe reaches, otherwise $Y_k^s(i)=0$.
\subsection{Sufficient Statistic}

We assume that each source sends probes independently, and the
observations of the arrivals of probes at nodes and receivers are also
assumed independent. Then, the loss process of a link is considered an
independent identical distributed ({\it i.i.d.}) process. This also
makes the collective impacts of probes sent by the sources to a link
{\it i.i.d.}. Further, the likelihood function of an experiment takes
either a product form of the individual likelihood function or a
summation form of the individual log-likelihood function. Therefore,
we will consider a single source first in the following discussion,
and then add a number of the single ones together to form a
multiple-sources general log-likelihood function.

If a probe sent by source $s$ reaches $Rs(i)$, the probe must pass
link $i$. Then, by examining the observations of $Rs(i)$, we are able
to confirm some of the probes sent by the source passing link $i$.
Based on the confirmed passes of each link, we have a set of
sufficient statistics to write the likelihood function of $\Omega$. To
obtain the confirmed passes of link $i$, we must examine each
observation received by $Rs(i), i\in E$. For each $i \in E$, let
\[
Y_{i}^s(j)=\max X_k^s(j), \mbox{     } k \in Rs(i).\] \noindent If
$Y_{c(i)}^s(j)=1$ probe $j$ reaches at least one of $Rs(i)$, that also
implies it reaching node $i$. Further, considering $Y_{i}^s(j)$ and
$Y_{p(s,i)}^s(j)$, we can confirm some of the probes sent by $s$
passing link $i$. If
\begin{enumerate}
\item  $Y_{i}^s(j)=Y_{p(s,i)}^s(j)=1$, it is confirmed that probe $j$ passes link
$i$; or
\item  $Y_{i}^s(j)=0$ and $Y_{p(s, i)}^s(j)=1$, it is confirmed that probe $j$ passes link  $p(s,i)$ but uncertain whether the probe passes link $i$; or
\item $Y_{i}^s(j)=Y_{p(s,i)}^s(j)=0$, it is obvious since uncertainty is transferable from a node to its descendants.
\end{enumerate}

\noindent If the first scenario occurs, we need to have $(1-\theta_i)$
in the likelihood function. If the second one occurs, we need to have
$(\theta_i + (1-\theta_i)(1-\beta_i))$ in the likelihood function,
where  $(1-\beta_i)=P(\bigwedge_{r\in R(i)} X_r =0|X_i=1;\theta)$ is
the loss rate of subtree $i$ that considers all possible combinations
that lead to $\bigwedge_{r \in R(i)} X_r =0$ since uncertainty is
transferable. If the last one occurs, we do not need adding any thing
into the likelihood function because its contribution to the
likelihood function is considered by one of its ancestors that has the
second scenarios for this probe.

Considering all probes sent by source $s$, and let
\[ n_i(s, 1)=\sum_{j=1}^{n^s} Y_{i}^s(j), \] \noindent denote the
number of the confirmed passes of the probes sent by source $s$ on
link $i$. In addition, let
\[ n_i(s, 0)=n_{p(s,i)}(s,1)-n_i(s, 1)\]

\noindent be the number of probes sent by source $s$ that turn to
uncertain in $T(i)$. Further, considering all sources, we have
$$n_{i}(1)=\sum_{s \in S(i)}n_{i}(s,1)\hspace{1cm}n_{i}(0)=\sum_{s \in S(i)}n_{i}(s, 0)$$
\noindent be the total number of probes confirmed from observations
that pass link $i$ and the total number of probes that turn to
uncertain in $T(i)$. $n_i(1), i \in E$ consist of a set of sufficient
statistics \cite{ZD09}.

\section{ Maximum Likelihood Estimator}
The statistics presented in the previous section are extended from
those presented in \cite{ZD09}. In \cite{ZD09}, we also attempted to
tackle the general topology, but did not derive a closed from
solution. Instead, we came up with a strategy to decompose a general
topology into a number of independent subtrees. Nevertheless, that
experience inspires us to further investigate the correlation in
$\Omega$, where an insight is discovered from this investigation. For
completeness, the link-based estimator is presented first. Then, we
present the insight which finally leads to a path-based estimator.
\subsection{Link-based Estimator}

Given the set of sufficient statistics, we can write the
log-likelihood function of $\Omega$. Using the same strategy as
\cite{Zhu06}, the log-likelihood function of a general network is
presented as follows:
\begin{eqnarray}\label{logL-mutisource}
L(\theta) &=&\sum_{i\in{E}}\Big[n_{i}(1)\cdot
log(1-\theta_{i})+n_{i}(0)\cdot
         log\xi_{i}\Big] \label{general tree}
\end{eqnarray}
\noindent where $\xi_i = \theta_i+(1-\theta_i)(1-\beta_i))$.
Differentiating $L(\theta)$ with respect to ({\it wrt.}) $\theta_i$
and setting the derivatives to 0, we have a set of equations. Solving
them, we have the followings:

\begin{eqnarray}
\theta_i=\left\{
  \begin{array}{l}
  1-\dfrac{\dfrac{n_i(S(i), 1)}{n^i}}{\beta_i},   \mbox{\hspace{2cm}       } i\in RL, \\
  1-\dfrac{\dfrac{ n_i(S(i),1)}{n_{f(i)}(S(i),1)+imp(S(i),f(i))}}{\beta_i}, \mbox{} i \in SBRL, \\
  1-\dfrac{\dfrac{\sum_{j \in S(i)} n_i(j,1)}{\sum_{j
\in S(i)}[n_{f(i)}(j, 1)+imp(j, f(i))]}}{\beta_i}, \\   \mbox{\hspace{4cm}} i\in SSNL, \\
\dfrac{\sum_{j \in S(i)}[ n_i(j,0)+ imp(j, f(i))]}{\sum_{j \in
S(i)}[n_{f(i)}(j, 1)+imp(j,f(i))]},  \mbox{\hspace{0.1cm}     } i \in
AOL,
\end{array}\right. \label{mgeneral}
\end{eqnarray}

\noindent where RL denotes root links, SBRL denotes the links that are
not root link but only receive probes from a single source, SSNL
denotes the links in intersections but not leave, and AOL denotes all
others, i.e. leaf links. Differentiating $L(\theta)$ {\it wrt.}
$\theta_i$ and setting the derivatives to 0. Then, we have a set of
equations as follows, one for a group of links. Let  $\beta_i =1, i
\in AOL$ and $imp(j, i)$ denotes the impact of $n_k(j,0)$, $k \in
a(j,i)$,
 on the loss rate of link $i$.
\begin{eqnarray}
&&imp(j,i)= \nonumber \\
&&\sum_{k \in a(j,i)}\dfrac{n_k(j,0)\cdot pa_i(k)\cdot\xi_i*\prod_{\substack{l \in a(i) \\
l \geq k}}\prod_{q \in C_l \setminus l} \xi_q}{\theta_k
+(1-\theta_k)\prod_{q \in C_{k}}\xi_q} \nonumber
\end{eqnarray}
\noindent where
\begin{eqnarray}
&&pa_i(k)=\prod_{\substack{l \in a(j,i) \\
l \geq k}} (1-\theta_l). \nonumber
\end{eqnarray}

Despite the estimate obtained by (\ref{mgeneral}) has been proved to
be the MLE in \cite{ZD09}, we have not found an analytical solution as
the one presented in \cite{ZD09} for the general topology since
(\ref{mgeneral}) is a transcendental equation in the form of
$\theta_i=f( \Theta)$, where $\Theta$ is the set of the parameters to
be estimated, one for a link in $E$. To overcome this, we proposed a
fixed-point algorithm to approximate the number of probes reaching a
joint node in \cite{ZD09}, and then a general network can be
decomposed into a number of independent trees in order to use the
estimator developed for the tree topology.

\subsection{Insight and Remark}

Despite there is no a closed form solution to (\ref{mgeneral}),  the
simple and uniform appearance of the four equations provides such an
insight of loss tomography that can be expressed in the following
remark:

{Remark:} regardless of the topology and the number of sources, the
MLE of the loss rate of a link can be obtained if we know:
\begin{enumerate} \item the total number of probes reaching the parent node of
the link, e.g. $n^i$, $n_{f(i)}(S(i), 1)+imp(S(i),f(i))$, or $\sum_{j
\in S(i)}[n_{f(i)}(j,1)+imp(j, f(i))]$ of (\ref{mgeneral});
\item the total number of probes reaching the receivers via
the link, e.g. $n_i(S(i),1)$,  or $\sum_{j \in S(i)} n_i(j,1)$ of
(\ref{mgeneral}); and,
\item the pass rate of the subtree rooted at the child node of the link, e.g. $\beta_i$ of (\ref{mgeneral}).
\end{enumerate}
Note that the link stated in the remark can be a path consisting of a
set of links serially connected. For the tree topology, there is only
one source, 1) and 2) can be suppressed to the pass rate of the path
connecting the source node to an internal node. Then, a polynomial
formula as the one presented in \cite{CDHT99}, i.e.

\begin{equation}
H_i(A_i, \gamma)=1-\dfrac{\gamma_i}{A_i}-\prod_{j \in d_i}
(1-\dfrac{\gamma_j}{A_i})=0, \label{minc}
\end{equation}

\noindent is obtained that expresses MLE of the path from the source
to node $i$. In fact, (\ref{minc}) expresses $1-\beta_i$, the loss
rate of subtree $i$, in two different forms of function of $A_i$: the
product of the loss rates of the multicast subtrees rooted at node $i$
and $1-\dfrac{\gamma_i}{A_i}$. Then, how to solve a high degree
polynomial becomes the key obstacle blocking a closed form solution to
the tree topology. The problem has been solved by \cite{ZD09} where an
equivalent transformation is proposed that merges multiple descendants
into two virtual ones.

 Using the same strategy as \cite{ZD09} and the set of sufficient statistics, (\ref{mgeneral}) shows the
impact of multiple sources on a link, regardless the link to be
estimated is in an intersection or not. The result, on one hand, shows
that as the tree topology, the cumulative impact is built from
 $n_j(s,0), j \in a(s,i)$ to link $i$; on the other hand, it shows that the impact comes from multiple sources, i.e.
 $imp(j, i), j \in S(i)$, in a general topology that makes a closed form estimator almost impossible.
 This is because
 (\ref{mgeneral}) is a multivariate polynomial relating to all links in a general topology. To find a solution, we need to
 to solve a set of multivariate polynomials, one for a link. That is
 intractable at least at this moment except for using approximation.

\subsection{Path-based Estimator}

There has been no path-based likelihood function for the general
topology as far as we know. This is partially due to the lack of
sufficient statistics of a link or a path from observations. With the
help of the minimal sufficient statistics proposed in Section 2, we
are able to write a path-based likelihood function here, and then
derive a set of likelihood equations, and finally have a path-based
estimator that provides MLE for the general topology.

Because of the intractability of the link-based estimator, our
attention is switched to a path-based estimator although the number of
paths is the same as the number of links. The path-based estimator
differs from the link-based one, based on the remark, in the following
aspacts:

\begin{itemize}
\item we know the total number of probes sent by a source; and
\item we know the total number of probes reaching $Rs(i)$ and know the sources of the arrived
probes.
\end{itemize}
\noindent As the tree topology, $\beta_i$ is unknown and can be
explicitly expressed by the available information. The difference
between the estimator of the tree topology and that of the general one
is at the intersections of trees in the general topology, which makes
the trees intersected dependent. Then, we need to determine the
correlation between the trees and the number of paths involved in an
intersection.

 Let
\begin{equation}
 A(s,i)=\prod_{j \in a(s,i)}(1-\theta_j), \mbox{\hspace{0.5cm}} \forall i \in V, \mbox{ and } \forall s
 \in S(i)
 \label{theta2A}
 \end{equation}

 \noindent be the pass rate of
the path from source $s$ to node $i$. It is easy to prove that
(\ref{theta2A}) is the bijection, $\Gamma$, from $\Theta$ to $A$,
where $\Theta$ is the support space of $\{\theta_i, i \in E\}$ and $A$
is the support space of $\{A(s,i), i \in V\setminus S, s \in S(i)\}$.
The statistics, $n_i(s, 1), i \in V, s \in S(i)$, are still the
sufficient statistics for the path-based likelihood function since the
number of probes sent by $s$ and observed by $Rs(i)$ is the confirmed
number of probes passing the path connecting $s$ to $i$. Then, we have
the following theorem for the MLE of a path in the general topology.

\begin{theorem} \label{general MLE}
The likelihood equation describing the pass rate of a path connecting
source $s$ to node $i$ can be expressed as
\begin{eqnarray}
A(s,i)=\dfrac{\gamma_i(s)}{\beta_i}, \mbox{\hspace{1cm}} s \in S(i).
\label{pathsetpoly}
\end{eqnarray}
where the empirical $\hat\gamma_i(s)=\dfrac{n_i(s,1)}{n^s}$
\end{theorem}
\begin{IEEEproof}
Using the sufficient statistics, we can write the path-based
likelihood function as

\begin{eqnarray}
P(A(s,i))&=&\prod_{i \in V\setminus S}\prod_{s\in S(i)} \Big
[A(s,i)^{n_i(s,1)}(1-A(s,i)+ \nonumber
\\
&&A(s,i)(1-\beta_i))^{(n_s(1)-n_i(s,1))}\Big]\nonumber \\
&=&\prod_{i \in V\setminus S}\prod_{s\in
S(i)}\Big[A(s,i)^{n_i(s,1)}\times \nonumber
\\ &&(1-A(s,i)\beta_i)^{(n_s(1)-n_i(s,1))} \Big] \nonumber
\end{eqnarray}

\noindent Changing it to the log-likelihood function, we have

\begin{eqnarray}
&&L(P(A(s,i))=\sum_{i \in V\setminus S}\sum_{s \in S(i)} \Big[ n_i(s,1)\log A(s,i) \nonumber \\
&&+(n_s(1)-n_i(s,1))\log(1-A(s,i)\beta_i)\Big ] \label{pathlikely}.
\nonumber
\end{eqnarray}

\noindent Differentiating it wrt $A(s,i)$ and let the derivative be 0,
we have

\begin{eqnarray}
A(s,i)=\dfrac{\gamma_i(s)}{\beta_i}, \mbox{\hspace{1cm}} s \in S(i).
\nonumber
\end{eqnarray}
\end{IEEEproof}

(\ref{pathsetpoly}) is identical to that obtained for the tree
topology except for having a condition for $s \in S(i)$. We call it
the consistent condition because it states the necessity of a
consistent $\beta_i$ for those trees intersected at node $i$. This
condition states that an MLE estimator must consider all probes sent
from $S(i)$ to $i$. Compared with the link-based estimator, the
path-based estimator only considers the intersected paths during
estimation that substantially reduces the correlation involved in
estimation.
 To obtain the MLE $\hat A(s,i)$ from the set of equations defined by (\ref{pathsetpoly}), we need:
 \begin{itemize}
 \item to obtain a consistent $\beta_i$ for all $\gamma_i(s), s\in S(i)$ and to express $\beta_i$ in two different ways as
 the functions of $A(s,i)$,
 \item to link the two expressions by an equation, and to prove there is a unique solution to the equation.
 \end{itemize}
 \noindent The following two theorems address
them, respectively.
\begin{theorem} \label{jointnodetheorem}
Let $A(k,i)$ be the pass rate of the path connecting $k, k \in S(i)$
to $i, i \in V \setminus S$ in a network of the general topology.
There is a polynomial, $H(A(k, i), S(i))$, as follows to express the
estimate of $A(k,i)$.
\begin{eqnarray}
&&H(A(k, i), S(i)) = 1-\dfrac{\hat\gamma_i(k)}{A(k,i)} - \nonumber \\
&& \prod_{j \in d_i}(1-\dfrac{\hat\gamma_i(k)\sum_{s \in S(i)} n_j(s,
1)}{A(k, i)\cdot \sum_{s \in S(i)} n_i(s,1)})=0 \label{generalpoly}
\end{eqnarray}
where

\begin{equation}
\hat\gamma_i(k)=\dfrac{n_i(k, 1)}{n^k} \nonumber
\end{equation}

\end{theorem}

\begin{IEEEproof}
Assume  $k \in S(i)$ sending probes to node $i$. Based on the first
equation of (\ref{mgeneral}), we have

\begin{eqnarray}
A(k,i)\beta_i = \dfrac{n_i(k,1)}{n^k}=\hat\gamma_i(k),
\mbox{\hspace{0.5 cm} }k \in S(i) \nonumber  \label{2connect}
\end{eqnarray}

\noindent  If $|S(i)|>1$, there are more than one sources sending
probes to node $i$, the pass rates of two sources, $s$ and $k, s, k
\in S(i)$, to node $i$ are correlated that can be expressed as
\begin{eqnarray}
A(s,i)=A(k,i)\dfrac{\hat\gamma_i(s)}{\hat\gamma_i(k)},
\mbox{\hspace{0.5cm} } s, k \in S(i)  \label{2relation}
\end{eqnarray}

\noindent If there is only one source for node $i$, (\ref{2relation})
can be ignored. Let $n_i^*(1)$ be the total number of probes reaching
node $i$, we have

\begin{eqnarray}
n_i^*(1) = \dfrac{A(k, i)}{\hat\gamma_i(k)}\sum_{s \in S(i)} n^s
\hat\gamma_i(s) \nonumber \label{totalnum}
\end{eqnarray}

\noindent Then, we have two different ways to express $1-\beta_i$.
Using the loss rate of subtree $j$ rooted at node $i$, we have
\begin{eqnarray}
1-\beta_i=\prod_{j \in d_i} \big[1-\dfrac{\sum_{s \in S(i)} n_j(s,
1)}{n_i^*(1)}\big] \label{subloss}
 \end{eqnarray}

\noindent and using (\ref{pathsetpoly}), we have

 \begin{eqnarray}
  1-\beta_i
  &=& 1- \dfrac{\hat\gamma_i(k)}{A(k,i)} \nonumber
  \end{eqnarray}
  \noindent Connecting the two, we have
\begin{eqnarray}
1- \dfrac{\hat\gamma_i(k)}{A(k,i)}  &=&\prod_{j\in d_i}
  \Big(1-\dfrac{\hat\gamma_i(k) \cdot \sum_{s \in S(i)} n_j(s, 1)}{A(k,i)\cdot \sum_{s \in S(i)}
  n^s
\hat\gamma_i(s)}\Big) \nonumber  \label{pathrateeql}
  \end{eqnarray}

\noindent Except for $A(k, i)$, all others, e.g. $\hat\gamma_i(k),
n_j(s,1)$, are either known or estimable from observations. Thus, the
above equation is a polynomial of $A(k, i)$. Alternatively, using
(\ref{subloss}) to replace $\beta_i$ from (\ref{mgeneral}), we have
the same result.
\end{IEEEproof}
(\ref{generalpoly}) generalizes (\ref{minc}) that considers various
paths ending at different nodes, including those having $|S(i)|=1$ and
those having $|S(i)>1$. For node $i$ having $|S(i)|=1$,
(\ref{generalpoly}) degrades to (\ref{minc}). For $|S(i)|>1$, if $j\in
d_i$,
\begin{equation}
\dfrac{\sum_{s \in S(i)} n_j(s, 1)}{\sum_{s \in S(i)} n_i(s,1)}
\label{alphai}
\end{equation}
is the estimate of the pass rate of the link connecting node $i$ to
node $j$, where the numerator is the sum of the probes sent by $S(i)$
that reach $Rs(i)$ while the denominator is the sum of the probes sent
by $S(i)$ that reach $Rs(j), j \in d(i)$. Both are obtainable from
$\Omega$. The estimate is built on the arithmetic mean that considers
the contribution of all sources sending probes to the link.

 Solving (\ref{generalpoly}), we have $\widehat A(k, i)$.
Then, we can have $\widehat A(s, i), s \in S(i)\setminus k$ and
 \[
\hat\beta_i=\dfrac{n^s}{\widehat A(s,i)}, s \in S.
\]
To prove the uniqueness of (\ref{pathsetpoly}), we have the following
Lemma

\begin{lemma}\label{property of likelihood equation} Assume $c_{i}\in(0,1)$, if $\sum_{i}c_{i}>1$,
$x=\prod_{i}\Big[(1-c_{i})+c_{i}x\Big]$ has a unique solution in
$(0,1)$. Otherwise, if $\sum_{i}c_{i}<1$, there is  either no solution
or have multiple solutions in $(0,1)$ for the equation.
\end{lemma}

\begin{IEEEproof}
See appendix \end{IEEEproof}

Based on Lemma \ref{property of likelihood equation}, we have
\begin{theorem} \label{unique theorem}
The set of likelihood equations formed by (\ref{pathsetpoly}) has the
unique solution.
\end{theorem}
\begin{IEEEproof}
Using $\dfrac{\sum_{s \in S(i)} n_j(s, 1)}{n_i^*(1)}$ to replace $c_i$
in Lemma \ref{property of likelihood equation}, the theorem follows.
\end{IEEEproof}

 To prove the estimate obtained by (\ref{generalpoly}) is the MLE, we
resort on a well known theorem for the MLE of a likelihood function
yielding the exponential family.
\begin{theorem}\label{properties of statistics}
If a likelihood function  belongs to a standard exponential family
with $A(s,i)$ as the natural parameters, we have the following
results:
\begin{enumerate}
  \item the likelihood equation $\frac{\partial
L(\theta)}{\partial \theta_i}=0$ has at most one solution
$\theta_i^*\in \Theta$;
  \item if $\theta_i^*$ exists, $\theta_i^*$ is the MLE.
\end{enumerate}
\end{theorem}
\begin{IEEEproof}
This theorem can be found from a classic book focusing on exponential
families, such as \cite{Brown86}. The likelihood function presented in
(\ref{pathlikely}) belongs to the exponential family, where $A(s,i)$
is the natural parameters. Then, the estimate obtained by
(\ref{generalpoly}) is unique in its support space and the MLE of
$A(s,i)$.
\end{IEEEproof}

\section{Solutions}

Given Theorem \ref{jointnodetheorem}, there are a number of ways  to
complete the estimation. One of them is to repeatedly use Theorem
\ref{jointnodetheorem} to obtain the pass rates of all paths, and then
use $\Gamma^{-1}$, from path rate to link rate, to have the loss rate
of each link, i.e.

\begin{equation}
1-\theta_i=\dfrac{\sum_{s \in S(i)} n^s A(s, i)}{\sum_{s \in S(i)} n^s
A(s, f(i))} \end{equation} The advantages of this is its uniformity,
while the disadvantage could be the requirement of solving a large
number of polynomials, some of them may be  high degree polynomials.

\subsection{Closed Form Solution}

It is easy to see the degree of (\ref{generalpoly}) is one less than
the number of descendants connected to node $i$ and we also know that
there is no closed form solution for a polynomial that has a degree of
5 or greater. Then, the key to have a closed form solution to
(\ref{generalpoly}) is to replace a multi-descendant node, i.e.  a
node with more than 5 descendants, with a tree. In \cite{ZD09}, such a
replacement is proposed for the tree topology and called a statistical
equivalent replacement since \begin{itemize} \item it is equal to use
a number of serially connected links to replace an egress link of a
multi-descendant node, and \item the statistics of the links
connecting or connected to the node remain the same as that before the
replacement. \end{itemize}. Comparing (\ref{minc}) with
(\ref{generalpoly}), one can notice that (\ref{minc}) is a special
case of (\ref{generalpoly}). Then, the replacement proposed for the
tree topology must be a special case of the general topology, where
the key rests on how to determine the statistics of the nodes located
in the tree used to replace a multi-descendant node. The following
theorem extends the result presented in \cite{ZD09} and provides the
answer to the above question.

\begin{theorem} \label{virtual link}
A multi-descendant node, say $i$,  can be replaced by a tree, where
the statistics of the nodes in the tree are determined according to
the observation of $R(i)$.  The following equation is used for an
introduced link $z$:

\begin{eqnarray}
&&n_z(1)=\sum_{s \in S(i)}\sum_{j \in d_z} n_j(s,1)-\sum_{s \in S(i)}\sum_{\substack { i<j \\
i, j \in d_z}} n_{ij}(s,1) + \nonumber \\ && \sum_{s \in S(i)}\sum_{\substack { i<j<k \\
i, j,k
\in d_z}} n_{ijk}(s,1)- \cdot\cdot +(-1)^{|d_z|-1} \sum_{s \in S(i)}n_{d_z}(s,1) \nonumber \\
\end{eqnarray}

\noindent where $d_z$ denotes the descendants connected to link $z$,
$n_{ij}(s,1) = \sum_{u=1}^n (Y_i^u (s)\wedge Y_j^u(s)) $ is the number
of probes that have been observed by at least one of R(i) and one of
R(j) simultaneously, similarly $n_{ijk}(s,1)=\sum_{u=1}^n (Y_i^u(s)
\wedge Y_j^u(s)\wedge Y_k^u(s)), \cdot\cdot\cdot $, and $ n_{d_z}(s,1)
= \sum_{u=1}^n (\bigwedge_{j \in d_z} Y_j^u(s)) $.
\end{theorem}

\begin{proof}
see appendix. \end{proof}

 Given
Theorem \ref{virtual link}, node $i$ can be replaced by a binary tree
when we estimate the pass rate of the paths connecting $S(i)$ to node
$i$. Then, equation (\ref{generalpoly}) becomes a linear equation of
$A(k, i), k \in S(i)$ and a closed form solution follows.

\subsection{Decomposition}

Apart from the closed form solution as above, another two closed form
solutions based on divide-and-conquer strategies are proposed as
alternatives. This is because equation (\ref{generalpoly}) shows such
a fact that the parameter estimation in a hierarchical structure with
the probes flowing in one direction can be carried out by dividing the
structure into a few segments as the d-separation presented in
\cite{JP00}.  For the general topology, given the number of probes
reaching a node, the descendants of the node become independent from
each other and can be estimated independently.  Then, the immediate
issue is which nodes should be selected as the decomposing points,
where the criterion used to evaluate the strategies is based on the
number of independent trees created after a decomposition, the smaller
the better since the computation cost of estimation is proportional to
the number of independent trees. The following theorem is presented
for the optimal strategy to select decomposing points.
\begin{theorem}
Given a general network covered by multiple trees, the optimal
strategy to decompose the overlapped multiple trees into a number of
independent trees is to use the nodes of $J$ as decomposing points.
\end{theorem}
\begin{IEEEproof}
We prove this by contradiction. Firstly, we assume there were two
strategies, say A and B, where A decomposes a general network into a
number of independent subtrees and there is at least one of the
decomposing points that is not a joint point and the total number of
independent subtrees is $m'$, while B decomposes the network at the
joint points only and the total number of independent subtrees is $m$.
If strategy A were better than strategy B,  we should have $m' < m$.
However, this is impossible: given the fact that all of the subtrees
rooted at a joint point can only become independent from each other if
we know the state of the joint point according to d-separation. Then,
we have $m' \geq m+1$, which contradicts to the assumption.
\end{IEEEproof}

 Given $i \in J$ and $n^*_i(1)$ obtained by (\ref{generalpoly}), a network of the general topology can be divided
 into a number of independent trees.
 The independent trees can be divided into two groups on the basis
 of the position of a tree relative to the decomposing point that separates it from others. We call the two groups:
 the ancestor group and the descendant one.
For the descendant group, given $n^*_i(1)$, the subtrees rooted at
node $i$ become independent from their ancestors and from each other.
Then, the loss rates of the links in the subtrees can be estimated by
assuming a source attached to the root that sends $n^*_i(1)$ probes to
the receivers. Further, the maximum likelihood estimators developed
for the tree topology, such as the top down algorithm \cite{Zhu06},
can be applied to each of the subtrees to estimate the loss rate of
each link of the subtree.

For the  ancestor group, the subtrees are also independent from each
other given $n^*_i(1)$. However, given $n^*_i(1)$ but the observations
at node $i$, we need to either use the first equation of
(\ref{mgeneral}) or the property of $imp(j, i)$ to estimate the loss
rates. The former leads to a modified estimator of (\ref{minc}). For
$s, s \in S(i)$, we use $\hat A(s,i)$ to replace $\gamma_i(s)$ in the
modified path-based estimator that has the following form:

\begin{eqnarray}
1-\dfrac{\gamma_{f^s(i)}(s)}{A(s, f^s(i))}&= &\big[1-\dfrac{\hat A(s,
i)}{A(s, f^s(i))}\big] \nonumber \\
&&\prod_{j \in d_{f^s(i)} \setminus
i}(1-\dfrac{\gamma_j(s)}{A(s,f^s(i))} ). \label{rateestimator}
\end{eqnarray}

\noindent Here, we also need to solve a polynomial to have $A(s,
f^s(i))$. Once having $A(s, f^s(i))$, we are able to estimate
$n^*_{f^s(i)}(1)$, then we move a level up and use
(\ref{rateestimator}) to estimate $A(s, f^s_2(i))$. This process
continues from bottom up until reaching $s$. If
 the tree being estimated has more than one intersections, at
the common ancestors of the intersections, the RHS of
(\ref{rateestimator}) will have a number of the left-most terms, one
for an intersection plus the product term for those subtrees that do
not link to any intersection. If the total number of overlapped trees
plus the number of independent subtrees is over 5, there is no direct
solution to (\ref{rateestimator}). Compared to the approach of using
(\ref{generalpoly}), this approach does not have significant advantage
since it also needs to solve a polynomial for a link. This shows the
path-based estimators are similar to each other in terms of
performance regardless of the topologies.

The question that we are facing is whether there is a more effective
approach than the path-based estimators and whether the link-based
estimators are better than the path-based one in this circumstance.
The answer is affirmative and the solution is reported in
\cite{Zhu11}.

\subsection{Data Consistency}
\label{consistent}

Data consistency  was raised in \cite{CDHT99} for three extreme
circumstances in observation $\Omega$ for the tree topology, which are
$\hat\gamma_i=0$, $1-\hat\theta_i>1$, and $\hat\gamma_k=\sum_{j \in
d_k} \hat\gamma_j$. The three extreme circumstances make the
estimation of $\theta_i$ an impossible task. To continue the
estimation without $\theta_i$, three corresponding procedures were
proposed to skip the impossible task. For the general topology, the
three circumstances still affect the estimation of the links, the
paths, and the subtrees that only observe the probes sent by a single
source, and the procedures proposed previously are still applicable to
the occurrences of the problems. However, due the the use of
multi-sources sending probes in a general topology, the situation is
different from the tree topology. We here specify the data consistent
issues for the paths and links that can observe probes sent by more
than one sources:
\begin{enumerate}
\item if $\gamma_i(s)=0, s \in S$, based on the location of link $i$, we need to
consider:
\begin{itemize}
\item if $i \in SBRL$, we remove $T(i)$ except for those that intersects with other trees. The loss rates of those links
located in an intersection can still be estimated as long as the
probes sent by other sources can reach the receivers attached to the
intersection.
\item if $i \in SSNL$, no link will be  deleted from the
subtree rooted at node $i$ as long as $Rs(i), s \in S(i)$ observes
some probes sent by the sources in $S(i)$. The loss rates of the links
in $T(i)$ cannot be estimated if no receiver in $Rs(i)$ observes a
probe from the sources.
\end{itemize}
\item as \cite{CDHT99} stated, $1-\hat\theta_i >1$ is not a feasible value. An estimator must avoid this circumstance in estimation.
The top down algorithm proposed in \cite{Zhu06} and the algorithm
based on (\ref{bottomup}) ensure this would not occur by having $\hat
n_{f(i)}(1) \geq \hat n_i(1), \forall i, i \in V$.
\item $\hat\gamma_i=\sum_{j \in d_i} \hat\gamma_j$ was raised in
\cite{CDHT99}. We call it partition circumstance here because the
equation shows that the observations of $R(j), j \in d_i $ are the
partitions of the observations of $R(i)$, i.e. the receivers attached
to each subtree of $T(i)$ has the exclusive observations for a portion
of the probes sent by the source. Given the partition circumstance at
node $i$ of a tree topology, equation (\ref{minc}) cannot be used to
estimate the loss rate of the links connecting $k$ to its children.
For the general topology, the situation is improved if $k$ is a parent
of node $j, j \in SSNL$. Even $\hat\gamma_k(s)=\sum_{j \in d_k}
\hat\gamma_j(s)$, we are still able to estimate the loss rate of the
link connecting $k$ to $j$ by (\ref{generalpoly}) as long as the
partition circumstance does not occur simultaneously for all $j$'s
parents.
\item \label{dataconsistent} It is clear that (\ref{minc}) is a special case of (\ref{generalpoly}).
Each of the three circumstances raised previously for data consistency
has its counterpart in the general topology that can be expressed as:
\begin{enumerate}
\item if $\sum_{s \in S(i)} \gamma_i(s)=0$,
\item if $1-\theta_i>1$, and
\item if $\sum_{s \in S(i)} n_i(s, 1) = \sum_{j \in d_i}
\sum_{s \in S(j)} n_j(s,1)$.  \end{enumerate} For the last one,
$A(s,i), s\in S(i)$ cannot be determined by (\ref{generalpoly}). If
one of the three scenarios occurs, the corresponding procedure
proposed in \cite{CDHT99} can be used.
\end{enumerate}

As discussed, data consistency has been improved in some degree for
the general topology. The improvement is due to the use of multiple
sources to send probes that reduces the probability of those extreme
circumstances in comparison with the tree topology.
\section{Statistical Property of the Estimators}

Apart from proving the three estimators are MLEs, we also study the
statistical properties of the estimators, such as whether the
estimators are minimum-variance unbiased estimators (MVUE); and/or the
estimate obtained is the best asymptotically normal estimates (BANE),
etc. This section provides the results obtained from the study.
\subsection{ Minimum-Variance Unbiased Estimator}
The estimators proposed in this paper are MVUE and the following
theorem proves this.
\begin{theorem} \label{MVUE theorem}
The estimators proposed in this paper are MVUE and the variances of
the estimates reach the Carm\'{e}r-Rao bound.
\end{theorem}
\begin{IEEEproof}
The proof is based on Rao-Blackwell Theorem that states that if g(X)
is any kind of estimator of a parameter $\theta$, then the conditional
expectation of g(X) given T(X), where T is a sufficient statistics, is
typically a better estimator of $\theta$, and is never worse. Further,
if the estimator is the only unbiased estimator, then, the estimator
is the MVUE.

To prove the estimator is an unbiased estimator, we only consider the
estimate of the binary tree since a multi-descendant tree can be
transformed into a binary one and equation (\ref{generalpoly}) is an
extension of the tree one. For the binary tree, the pass rate of link
$i$, $A_i$, is estimated by
\[
\dfrac{\hat\gamma_1
\hat\gamma_2}{\hat\gamma_1+\hat\gamma_2-\hat\gamma_i}=\dfrac{\hat\gamma_1
\hat\gamma_2}{\hat\gamma_{12}}= \dfrac{n_1(1)n_2(1)}{n*n_{12}(1)}
\]
where $\hat\gamma_{12}=\dfrac{n_{12}(1)}{n}$ is the empirical pass
rate of the probes that reach both node 1 and node 2. Then, we have
\begin{eqnarray}
&&E\Big (\dfrac{n_1(1)n_2(1)}{n*n_{12}(1)} -  A_i\Big ) \nonumber \\
&=& E\Big (\dfrac{n_1(1)n_2(1)}{n*n_{12}(1)}\Big ) - E(A_i) \nonumber \\
&=& E\Big (\dfrac{n_1(1)n_2(1)}{n_i(1)n_{12}(1)}\cdot \dfrac{n_i(1)}{n} \Big ) -\dfrac{1}{n}E(\sum_{k=1}^n X_i^k)\nonumber \\
&=&\dfrac{1}{n}E\Big ( \dfrac{\sum_{k=1}^{n_i(1)} X_1^k \cdot
\sum_{k=1}^{n_i(1)} X_2^k}{n_i(1)\sum_{k=1}^{n_i(1)} X_{12}^k} \cdot
\sum_{k=1}^n X_i^k \Big )-
\dfrac{1}{n}E(\sum_{k=1}^n X_i^k) \nonumber \\
&=&\dfrac{1}{n}E\Big ( \dfrac{\sum_{k=1}^{n_i(1)} X_1^k \cdot
\sum_{k=1}^{n_i(1)} X_2^k}{n_i(1)\sum_{k=1}^{n_i(1)} X_{12}^k}  \Big )
\cdot E(\sum_{k=1}^n X_i^k )- \nonumber \\
&& \dfrac{1}{n}E(\sum_{k=1}^n X_i^k) \nonumber \\
&=&\dfrac{1}{n}E(\sum_{k=1}^n X_i^k)-\dfrac{1}{n}E(\sum_{k=1}^n
X_i^k)=0
\end{eqnarray}
The statistics used in this paper has been proved to be the minimal
complete sufficient statistics. Based on Theorem \ref{unique theorem},
we have a unique solution in $(0,1)$ for the parameter. Then, applying
Rao-Blackwell theorem, the theorem follows.

Given theorem \ref{MVUE theorem}, it is easy to prove the variance of
the estimates, e.g. $\hat\theta_i$ and $\hat \beta_i$, obtained by
(\ref{generalpoly}) are equal to Carm\'{e}r-Rao low bound since
(\ref{pathlikely}), the likelihood function leading to
(\ref{generalpoly}), belongs to the standard exponential family.
\end{IEEEproof}
Based on Fisher information we can prove the variance of the estimates
obtained by (\ref{generalpoly}) from $\Omega$ is also smaller than
that of an estimate obtained by (\ref{minc}) from $\Omega_s, s \in S$.
Since the receivers attached to intersections observe the probes sent
from different sources, the sum of the observed probes is at least
larger than or equal to the maximum number of probes observed from a
single source. Therefore, there is more information about the loss
rates of the links located in the intersections since information is
addictive under {\it i.i.d.} assumption. With more information, the
variance of an estimate of a parameter must be smaller than another
obtained from the probes sent by a single source according to Fisher
information. Given the less varied estimates from the intersections,
the variances of the estimates of other links that are not in
intersections are also reduced, at least not increased. Therefore, the
estimates obtained by (\ref{generalpoly}) is better than those
obtained from a single source.

\subsection{General Topology vs. Tree Topology}

The results presented in this paper can be viewed as a generalization
of the results presented in \cite{CDHT99} since the tree topology is
only a special case of the general topology. The findings and
discovery presented in this paper cover those presented in
\cite{CDHT99}. The following corollary confirms this:
\begin{corollary}\label{special case}
Any discovery, including theorems and algorithms, for loss estimation
in the general topology, holds for the tree topology as well.
\end{corollary}

For instance, (\ref{generalpoly}) obtained for the general topology
holds for the tree one. When $S(i)=\{k\}$,  we have
\begin{eqnarray}
&&H(A(s, k), S(k)) = 1-\dfrac{\gamma_i(k)}{A(k,i)} - \nonumber \\ &&
\prod_{j \in d_i}(1-\dfrac{\gamma_i(k)\sum_{s \in S(i)} n_j(s,
1)}{A(k, i)\cdot \sum_{s \in S(i)} n_i(s,1)}) \nonumber \\
&=&1-\dfrac{\gamma_i(k)}{A(k,i)} -\prod_{j \in
d_k}(1-\dfrac{\gamma_i(k) n_j(1)}{A(k,i)\cdot n_i(1)}) \nonumber \\
&=&1-\dfrac{\gamma_i(k)}{A(k,i)} -\prod_{j \in
d_k}(1-\dfrac{\gamma_j(k)}{A(k,i)})=0
\end{eqnarray}

\noindent the last equation is  $H(A_i, i)$ presented in
\cite{CDHT99}. Another example has been presented in Section
\ref{consistent} \ref{dataconsistent} for data consistency.

 On the other hand, we are also interested in whether
those properties discovered from the tree topology can be extended
into the general one and the difference between the original
properties and the extended ones, in particular for the rates of
convergence.

\subsection{Large Sample Behavior of the Estimator}

Since the estimate obtained by the proposed estimator is the MLE $\hat
\theta_i$, we are able to apply some general results on the asymptotic
properties of MLEs in order to show that $\sqrt n(\hat \theta_i -
\theta_i)$ is asymptotically normally distributed as $n\rightarrow
\infty$. Using this, we can estimate the number of probes required to
have an estimate with a given accuracy for many applications. The
fundamental object controlling convergence rates of the MLE is the
Fisher Information Matrix at $\theta_i$. Since $L(\theta)$ yields
exponential family, it is straightforward to verify that $\hat
\theta_i$ is consistent and that $L(\theta)$ satisfies conditions
under which $\textit{I}$ is equal to

\[
I_{jk}(\theta)=-E\dfrac{\partial^2L}{\partial \theta_j \partial
\theta_k}(\theta)
\]

\noindent Eliminate singular on the boundary of $(0,1]^{|E|}$, we have
\begin{theorem}\label{asympototic theorem}
When $\theta_i \in (0,1), i \in V\setminus S, \sqrt n(\hat \theta_i -
\theta_i)$ converges in distribution as $n\rightarrow \infty$ to an
$|E|$ dimensional Gaussian random variable with mean 0 and covariance
matrix $I^{-1}(\theta)$, i.e.
\[
\sqrt n (\hat\theta -\theta) \xrightarrow{D} N(0,I^{-1}(\theta))
\]

\noindent and $\hat \theta_i$ is the best asymptotically normal
estimate (BANE).
\end{theorem}

\begin{IEEEproof}
\noindent It is know that under following regularity conditions:
\begin{itemize}
\item the first and second derivatives of the log-likelihood function
must be defined.
\item the Fisher information matrix must not be zero, and must be
continuous as a function of the parameter.
\item the maximum likelihood estimator is consistent.
\end{itemize}
\noindent the MLE has the characteristics of asymptotically optimal,
i.e., asymptotically unbiased, asymptotically efficient, and
asymptotically normal. The characteristics are also called BANE.

It is clear that (\ref{pathlikely}), the likelihood function used in
this paper, belongs to the standard exponential family, which ensures
the consistence and uniqueness of the MLE. To satisfy  the second
condition for the exponential family, $n_i(s, 1), i \in E, s \in S$
should not be linearly related, this is true as $n_i(s,1), i \in E, s
\in S$ have been proved to be the minimal sufficient statistics, see
Theorem 2 of \cite{ZD09}. Then, we only need to deal with the first
condition. Obviously, (\ref{mgeneral}) has both first and second
derivatives in $(0,1)^{|E|}$, and $L(\theta)$ is strictly concave,
which ensures the Fisher information matrix $I(\theta)$ positive
definiteness.
\end{IEEEproof}

Theorem  \ref{asympototic theorem} states such a fact that with the
increase of the number of probes sent from sources, there are more
probes reaching the links of interest. Then, there is more information
for the paths to be estimated. Let $I_0(\theta)$ is the Fisher
Information for a single observation, we have
$I(\theta)=nI_0(\theta)$, where $n$ is the number of observations
related to the link/path being estimated.

As $n\rightarrow \infty$, the difference in terms of Fisher
information between the two estimators approaches to zero. Therefore,
as $n \rightarrow \infty$, estimation can be carried out on the basis
of individual tree and the asymptotical properties obtained previously
for the tree topology \cite{CDHT99} hold for the general topology as
well. On the other hand, if $n<\infty$, the estimate obtained by
(\ref{generalpoly}) is more accurate than those obtained from an
individual tree. The simulation results presented in the next section
illustrate this that shows the fast convergence of the estimates for
the links located in the intersection because there are more
information about the links.

 Although there are a number of large sample properties that can
be related to loss inference in the general topology, including
various asymptotic properties, we would not discuss them further
because $n \rightarrow \infty$ means infinite number of probes be sent
from sources to receivers that requires a long period of stationarity
of the network, including traffic and connectivity, which is
impractical based on the measurement \cite{PAX99}.

\section{Simulation Study}

\begin{figure}
\centerline{\psfig{figure=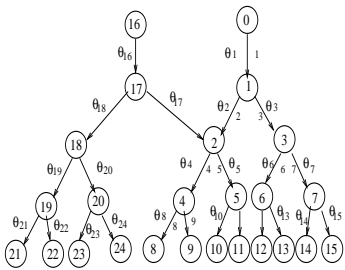,height=3.5cm,width=4.5cm}}
\caption{A Network with Multiple Sources} \label{multisource}
\end{figure}

\begin{figure}
\begin{center}
\epsfig{figure=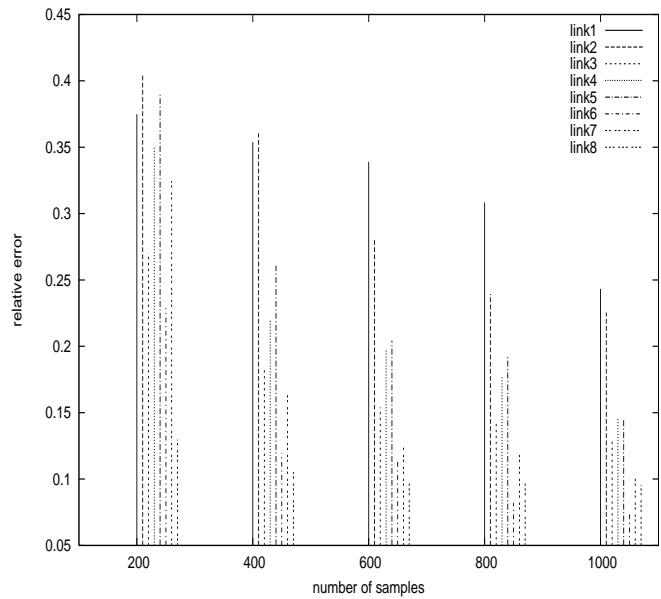,height=9.0cm,width=8.0cm,angle=-90}
\caption{Relative Error for Link 1-8} \label{rel1-8}
\end{center}
\end{figure}

\begin{figure}
\begin{center}
\epsfig{figure=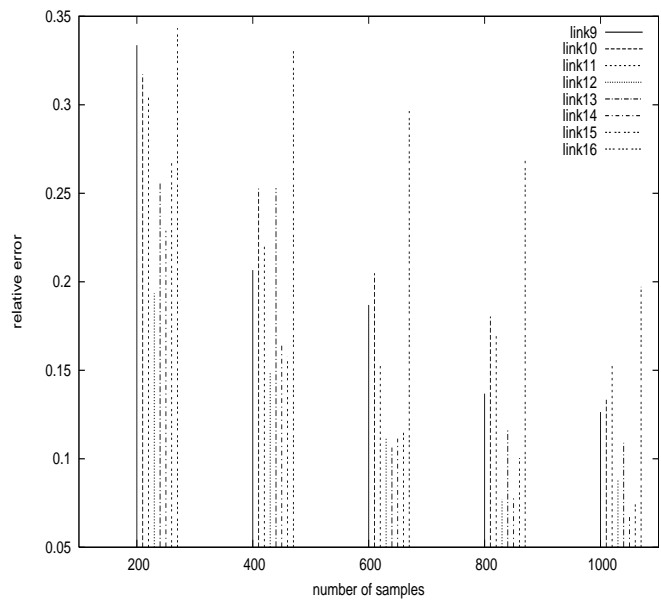,height=9.0cm,width=8.0cm,angle=-90}
 \caption{Relative Error for Link 9-16}
\label{rel9-16}
\end{center}
\end{figure}

\begin{figure}
\begin{center}
\epsfig{figure=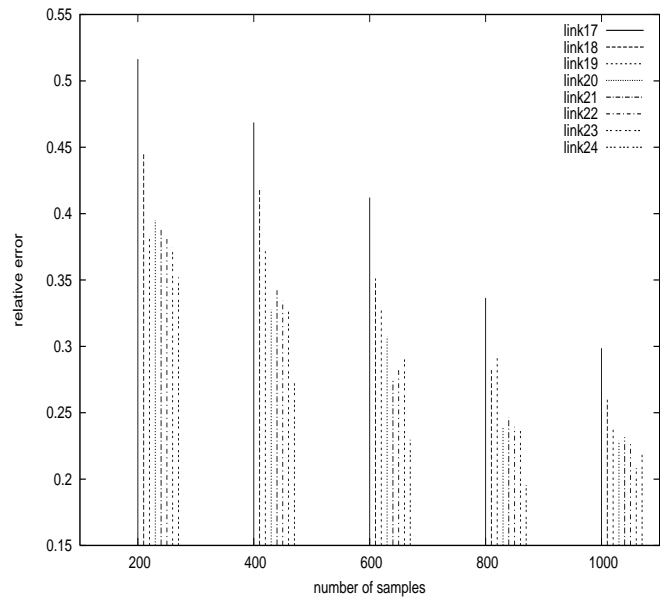,height=9.0cm,width=8.0cm,angle=-90}
 \caption{Relative Error for Link 17-24}
\label{rel17-24}
\end{center}
\end{figure}

For the purpose of proof of concept, a series of simulations were
conducted on a simulation environment built on {\it ns2}, the network
simulator 2 \cite{NS2}, the network topology used in the simulation is
shown in Figure \ref{multisource}, where two sources located at node 0
and node 16
 multicast probes  to the
attached receivers in an interval of 0.01 second. The binary network
used here is for the simplicity reasons since no matter how many
branches a node has they are always converted to a two-layer binary
tree in estimation. Apart from the traffic created by probing, a
number of TCP sources with various window sizes and a number of UDP
sources with different burst rates and periods are added at the roots
and internal nodes to produce cross-traffic, where TCP traffic takes
about $80\%$ of the total traffic. The loss rate of each link is
controlled by a random process that has $1\%$ drop rate except link
$8$ that has $10\%$ drop rate. Each simulation run for 200 simulation
seconds, and executed under the same condition except for using
different seeds. The samples collected in this period are divided into
groups to study the effect of group size on the accuracy of
estimation. Five groups for 200, 400, 600, 800, and 1000 samples are
presented here  to illustrate the impact of the group sizes on the
accuracy of estimation. The accuracy is measured by the relative error
that is defined as:

\[
\dfrac{abs(\mbox{actual loss rate - estimated loss
rate})}{\mbox{actual loss rate}}
\]

\noindent where we use ns2 to report the losses and passes incurred on
individual links. The ratio between the losses and the sum of the
losses and passes is used as actual loss rate. The estimated loss rate
of a link is obtained by  only considering the end-to-end
observations. The resultant relative errors of the 24 links are
presented in Figure \ref{rel1-8} to Figure \ref{rel17-24}, each for 8
links. The figures show that in general with the increase of the group
size, the relative errors are decreased as expected. Although there
are a few cases of slight increase in the figures, that is due to the
randomness of relative small group sizes used in estimation. The
problem can be overcome with the increase of the group sizes. For
example, observing Figures \ref{rel1-8} and \ref{rel9-16}, one is able
to notice the consistent decrease of the relative errors with increase
of the group size for those links falling into the intersection, i.e.,
link 4, 5, and 8-11 because there are almost 2 times of probes
traversing through this intersections in comparison with other links.
The relative errors of the shared links drop below 15\% when each
source sends 1000 probes; when 2000 probes are sent by each source,
the relative errors of those links drop below 10\% (not presented
here). This phenomenon suggests the use of large group size in
estimation for high accuracy. However, whether we are able to use more
samples in estimation is a question that is not only related to the
complexity of an algorithm but also related to the time scale of the
stationarity of the network of interest. Based on the setup of our
simulation, a 20-second (0.01 $\times$ 2000) stability is needed to
have the relative error drops to 10\%. If the time scale of the
stability of a network is much larger than 20 seconds, the estimates
obtained by the proposed estimators have potential to be used in
traffic engineering to control traffic flows.

From Figures \ref{rel1-8} and \ref{rel9-16},  there is a noticeable
difference between link 8 and link 9 in their converging rate in terms
of the relative error. The two links are siblings, that means the same
number of probes observed by their common parent first before sending
to receiver 8 and 9. However, the relative error of link 8 converges
much quick than that of link 9, while link 8 has a much higher loss
rate than link 9 does. This indicates that the accuracy of estimating
the loss rate of a link is proportional to the number of probes
observed by its siblings. This can be explained analytically. For
instance, given a 2 layer binary tree with 4 nodes, node 0 is the root
that connects to node 1, node 1 has two children 2 and 3 and uses link
2 and 3 to connect them; the MLE of the loss rate of link 3 is equal
to

\[ \hat\theta_3=\dfrac{n_3(0)}{n_2(1)}
\]

\noindent \cite{Zhu06-1}. In this simulation, the loss rate of link 8
is set to 10\% that is 10 times of that of link 9, which makes
$n_8(1)$ much smaller than $n_9(1)$ and vise verse for $n_8(0)$ and
$n_9(0)$. This implies that we get more information for link 8 than
that for link 9 from observations. That is why the relative error of
link 9 is higher than that of link 8. This leads to a finding that
extends the finding of \cite{CDHT99} on branching ratio where the
authors stated that estimate variance reduces with the increase of
branching ratio. The fact of increasing branching ratio is equal to
the decrease of the loss rate of the siblings of a link being
estimated, that leads to more information about the link from the
observations of the siblings. Therefore, to have an accurate estimate
of a link, apart from increasing the number of samples in a group we
can increase the number of siblings of a link instead. In contrast to
previous algorithms, the increase of siblings would not significantly
affect the complexity of the algorithms proposed in this paper since
they all have $O(|E|)$ complexity.


\section{Conclusion}
In this paper, we presents our recent findings on loss tomography of
the general topology, including both theoretical results and practical
algorithms. In theory, we extend the set of complete minimal
sufficient statistics proposed previously for the link-based estimator
into the path-based estimator. Based on the statistics, we provide the
path-based likelihood function and have a set of likelihood equations.
Solving the likelihood equations, we derive a direct expression of the
MLE for the pass rate of the paths in a general topology. The direct
expression has a similar structure as its counterpart in the tree
topology, a polynomial is for a path. The expression in the general
topology differs from the one in the tree topology in the collective
constraint, where a set of polynomials is bounded together if the
paths end at the same node that has a descendant tree.  In this
aspect, the polynomial derived for the tree topology \cite{CDHT99} can
be regarded as a special case of the one derived in this paper.

With the direct expression, we are able to estimate the number of
probes reaching the root of an intersection. Given the estimated
number of probes reaching the roots of all intersections, the network
can be decomposed into a number of independent trees. For the
descendant trees, the algorithms proposed for the tree topology can be
applied to them to estimate the loss rates of the links. For the
ancestor trees that are not from the intersections, new algorithms are
needed since there is no exact observations for the leaf nodes that
are the joint nodes before decomposition. Two types of algorithms,
path-based and link-based, are presented and compared in this paper.
The result shows that the link-based one is significantly better than
the path-based one here in terms of performance.

The method proposed in \cite{ZD09} to merge subtrees rooted at the
same link has been extended into the general topology to merge the
statistics obtained from multiple sources. Using this method, we are
able to reduce the degree of the polynomials used to express the pass
rate of a path, and subsequently avoid the use of a numeric method to
solve a high degree of polynomial. The estimator proposed in this
paper consists of 3 steps

\begin{enumerate}
\item using (\ref{generalpoly}) to estimate the pass rates of the paths connecting sources to joint nodes,
\item using the top down algorithm to estimate the loss rates of the links in descendant
trees, and
\item using (\ref{bottomup}) to estimate the loss rate of the links in
ancestor trees. \end{enumerate}

If no complicated overlapping occurs between the trees used to cover a
network, there is no need to have any iterative procedures involved in
estimation. The complexity of the estimator is $O(|E|)$. This
performance is significantly better than the iterative algorithms,
including the EM.

Data consistency raised previously for the tree topology \cite{CDHT99}
has been reconsidered for the general topology. Our study shows this
problem remains but has been eased when multiple sources are used to
send probes to receivers. In addition, our study shows the convergence
rate of the links that receive probes from more sources is faster than
those that only receive from one.

\section*{Appendix}

\noindent{\bf Lemma \ref{property of likelihood equation}}
\begin{IEEEproof} Let $h_1(x)=x$ and $h_2(x)=\prod_i\Big[ (1-c_i)+c_i x \Big]$, we
have $h_1'(x)=1$ and $h_2'(x)=h_2(x)\sum_i\frac{c_i}{(1-c_i)+c_i x}$.
Let $q_i= \sum_i\frac{c_i}{(1-c_i)+c_i x}$, we have $h_1^{''}(x)=0$
and $h_2^{''}(x)=h_2(x)[(\sum_i q_i)^2 - \sum_i q_i^2]>0$, if $x \in
[0,1]$. Let $h(x)=h_1(x)-h_2(x)$, that is strictly concave on $[0,1]$.
In addition, $h^{'}(0)=1-\prod_i (1-c_i)\sum_i \frac{c_i}{1-c_i}>0$
and $h^{'}(1)=1-\sum_i c_i$. If $\sum_i c_i>1$, $h^{'}(1)<0$, there is
a unique solution to $h(x)=0$ for $x \in [0,1]$ since $h(x)$ is
continuously differentiable on $[0,1]$, and $h(0)=-\prod_i (1-c_i)<0$
and $h(1)=0$. Otherwise, if $\sum_i c_i<1$, $h^{'}(1)>0$, there is no
solution to $h(x)=0$ for $(0,1)$.
 \end{IEEEproof}

\noindent{\bf Theorem \ref{virtual link}}
\begin{IEEEproof}
It has been proved that each multicast tree or subtree can be treated
as a virtual link in estimation \cite{CDHT99}. Here, we want to prove
that a number of virtual links rooted at the same node can be
considered a virtual link in estimation if the statistics of the
subtrees are merged according to the theorem. To prove such a
transformation is an equivalent one, we need to prove the following
\begin{enumerate}
\item the statistics of node $i$ remains the same after the
transformation. \label{virtuala}
\item the estimate of $\theta_i$ obtained from the internal view of the transformation is equal to that from the internal views of the
origin. \label{virtualb}
\end{enumerate}

 It
is known that $n_i(s,1)=\sum_{l=1}^{n(s)} (\bigvee_{p\in d_i}
Y_p^l(s))$. If we divide $d_i$ into $d_{i1}$ and $d_{i2}$ and let
$Y_{d_{ik}}^l(s) =\bigvee_{p\in d_{ik}} Y_p^l(s), k \in \{1, 2\}$ be
the observation of $d_{ik}$ for probe $l$ sent by source $s$, $\sum_{s
\in S(i)} n_i(s, 1)=\sum_{s \in S(i)}\sum_{l=1}^{n(s)}
(Y_{d_{i1}}^l(s) \bigvee Y_{d_{i2}}^l(s))$ is the statistics of
$d_{ik}$. If $d_{ik}, k \in \{1, 2\} $ is merged into a virtual link
and node $ik$ is assumed to be the child node of the virtual link,
then

\begin{eqnarray}
 n_{ik}(1) &=& \sum_{s
\in S(i)}\sum_{l=1}^{n(s)} \bigvee_{j\in d_{ik}} Y_j^l(s) \nonumber \\
&=&\sum_{s \in S(i)}\Big [\sum_{j \in d_{ik}} n_j(s,1)-\sum_{\substack
{ i<j
\\ i, j \in d_{ik}}} n_{ij}(s,1) + \nonumber \\ && \sum_{\substack {
i<j<k \\ i, j,k \in d_{ik}}} n_{ijk}(s, 1)- \cdot\cdot
+(-1)^{|d_{ik}|-1} n_{d_{ik}}(s,1) \Big  ] \nonumber \label{expend}
\end{eqnarray}
\noindent Then, \ref{virtuala}) holds.

Given $n_{ik}(1), k \in \{1, 2\}$, the RHS of (\ref{pathrateeql}) can
be written as

\begin{eqnarray}
RHS&=& \Big(1-\dfrac{n_{i1}(1)}{A(k,i)\cdot \sum_{s \in S(i)} n_s(1)
\hat\gamma_i(s)}\Big) \cdot  \nonumber \\
&&\Big(1-\dfrac{n_{i2}(1)}{A(k,i)\cdot \sum_{s \in S(i)} n_s(1)
\hat\gamma_i(s)}\Big)
\end{eqnarray}

\noindent To avoid the partition problem, when we select subtrees to
merge, we need to ensure the observations of the two group receivers
are intersected. Then, the theorem follows.

\end{IEEEproof}

\bibliography{congestion}

\end{document}